\documentclass[final,5p,times,twocolumn]{elsarticle}
\usepackage[dvipdfmx]{graphicx}
\usepackage{amsmath}
\usepackage{color}
\def\vc#1{\mbox{\boldmath $#1$}}
\def\Be{{^{8}{\rm Be}}}
\def\C{{^{12}{\rm C}}}
\def\Ox{{^{16}{\rm O}}}

\def\LH{{^{5}_\Lambda{\rm He}}}
\def\LB{{^{9}_\Lambda{\rm Be}}}
\def\LC{{^{13}_{\ \Lambda}{\rm C}}}

\usepackage{amssymb}

\journal{Physics Letters B}

\begin{document}

\begin{frontmatter}



\title{Multi-cluster dynamics in $\LC$ and analogy to clustering in $\C$}


\author{Y. Funaki$^{a,b}$, M. Isaka$^c$, E. Hiyama$^b$, T. Yamada$^d$, and K. Ikeda$^b$}
\address{$^a$School of Physics and Nuclear Energy Engineering and IRCNPC, Beihang University, Beijing 100191, China \\
$^b$Nishina Center for Accelerator-Based Science, The Institute of Physical and Chemical Research (RIKEN), Wako 351-0198, Japan \\
$^c$Research Center for Nuclear Physics (RCNP), Osaka University,\\ Ibaraki 567-0047, Japan \\
$^d$Laboratory of Physics, Kanto Gakuin University, Yokohama 236-8501, Japan}

\begin{abstract}
We investigate structure of $\LC$ and discuss the difference and similarity between the structures of $\C$ and $\LC$ by answering the questions if the linear-chain and gaslike cluster states, which are proposed to appear in $\C$, survives, or new structure states appear or not. We introduce a microscopic cluster model called, Hyper-Tohsaki-Horiuchi-Schuck-R\"opke (H-THSR) wave function, which is an extended version of the THSR wave function so as to describe $\Lambda$ hypernuclei. We obtained two bound states and two resonance (quasi-bound) states for $J^\pi=0^+$ in $\LC$, corresponding to the four $0^+$ states in $\C$.
However, the inversion of level ordering between the spectra of $\C$ and $\LC$, i.e. that the $0_3^+$ and $0_4^+$ states in $\LC$ correspond to the $0_4^+$ and $0_3^+$ states in $\C$, respectively, is shown to occur. 
The additional $\Lambda$ particle reduces sizes of the $0_2^+$ and $0_3^+$ states in $\LC$ very much, but the shrinkage of the $0_4^+$ state is only a half of the other states, in spite of the fact that attractive $\Lambda$-$N$ interaction makes nucleus contracted so much when the $\Lambda$ particle occupies an $S$-orbit.
In conclusion, the Hoyle state becomes quite a compact object with $\LB+\alpha$ configuration in $\LC$ and is no more gaslike state composed of the $3\alpha$ clusters. Instead, the $0_4^+$ state in $\LC$, coming from the $\C(0_3^+)$ state, appears as a gaslike state composed of $\alpha+\alpha+\LH$ configuration, i.e. the Hoyle analog state. A linear-chain state in a $\Lambda$ hypernucleus is for the first time predicted to exist as the $0_3^+$ state in $\LC$ with more shrunk arrangement of the $3\alpha$ clusters along $z$-axis than the $3\alpha$ linear-chain configuration realized in the $\C(0_4^+)$ state. All the excited states are shown to appear around the corresponding cluster-decay threshold, reflecting the threshold rule.
\end{abstract}


\end{frontmatter}



Nuclear clustering, as well as the mean-field-type correlation, is an important basis to understand nuclear many-body systems~\cite{Wi77,Ho12}. Appearance of the clustering is strongly associated with cluster-decay thresholds. The so-called Ikeda diagram, which refers to the necessary condition about the excitation energy for the formation of cluster states, tells us that a cluster structure appears around the corresponding threshold, while if the excitation energy is low enough below the threshold, the cluster structure dissolves into a mean-field state as a result of strengthened interaction between the clusters~\cite{Ik68,Zh17}. This condition known as the threshold rule allows us to consider that if a certain system involves a variety of thresholds, a variety of corresponding cluster states likely to appear. 

\begin{figure}[htbp]
\begin{center}
\includegraphics[scale=0.83]{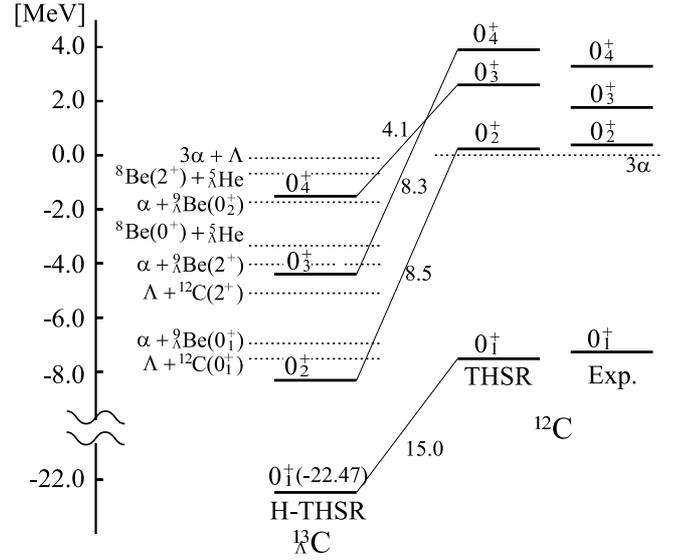}
\caption{(color online) Spectra of $\C$ and $\LC$ calculated with THSR and H-THSR wave functions. Experimental spectrum of $\C$ is also shown~\cite{It11}. Numbers, $15.0$, $8.5$, $8.3$, and $4.1$, are ${\cal E}_\Lambda$ values in a unit of MeV, which are defined as differences of binding energies of the states of $\LC$ from those of the corresponding states in $\C$ (see the text for more detailed definition of ${\cal E}_\Lambda$). Spectrum of $\C$ calculated with THSR wave function is taken from Refs.~\cite{Fu15,Fu16}.}
\label{fig:level}
\end{center}
\end{figure}

This threshold rule for nuclear clustering seems to be more important in light $\Lambda$ hypernuclei~\cite{Hi09}. For example, in $\C$, the lowest threshold is the one to decay into $3\alpha$ clusters located at $7.27$ MeV. The famous Hoyle state~\cite{Ho54,Co57,Fy11} exists at $0.38$ MeV above the threshold and is known to have well developed $3\alpha$ cluster structure. However, the situation amazingly changes if a $\Lambda$ particle joins in $\C$, namely being $\LC$. There appear new cluster species with the $\Lambda$ particle, i.e. $\LH$ and $\LB$, which are bound by $3.12$ MeV and $6.71$ MeV below the $\alpha+\Lambda$ and $2\alpha+\Lambda$ thresholds, respectively. The ground state of $\LC$ is further bound by $11.69$ MeV, relative to the $\C+\Lambda$ threshold. Therefore in $\LC$, the $\Be+\LH$ and $\alpha+\LB$ thresholds, as well as the $\C+\Lambda$ threshold, newly show up below the $3\alpha+\Lambda$ threshold (see Fig.~\ref{fig:level}). This situation implies that in $\LC$, according to the threshold rule, various cluster structures, which are not limited to the $3\alpha+\Lambda$ clustering, play an important role in the formation of its excited states.

In this study, we focus on the many-body cluster dynamics appearing in $\LC$, and investigate the structures of its excited states up to around the $3\alpha+\Lambda$ threshold. If the $\Lambda$ particle is put into a core nucleus, what kind of structure change is expected has been an important issue in structure studies of hypernuclei. The above argument may then give an important hint to answer this question.

Although in $\Lambda$ hypernuclei, two-body cluster systems like ${^7_\Lambda {\rm Li}}$, $\LB$, etc~\cite{Ya85,Hi97,MBI83,Ya84,YMB86,Hi99,Is11,Ba83,Ya85a}, have well been investigated, more-body cluster systems were only poorly discussed so far.
In $\LC$, which is obviously a typical example of multi-clustered hypernuclei, several authors have discussed its nature by using cluster models, like the Resonating Group Method (RGM)~\cite{Ya85,Wh37} and Orthogonality Condition Model (OCM)~\cite{Hi97,Sa68,Hi00}. Concerning the structure aspect, however, a special emphasis was only put on the difference of properties that appears when the $\Lambda$ particle couples with the shell-model-like state or the cluster state. They considered the $(1/2)_2^+$ state to correspond to the Hoyle analog state in which the $\Lambda$ particle couples with the Hoyle state. They found that the $(1/2)_2^+$ state has much smaller r.m.s. radius than that of the Hoyle state in $\C$, due to a gluelike role of the $\Lambda$ particle, while the ground state has negligibly small contraction by the $\Lambda$ particle, due to its saturation property. This is because the Hoyle state has very dilute density which is about 1/3 of saturation~\cite{Fu08a}, so that a much larger shrinkage effect was obtained when the $\Lambda$ particle is added to the Hoyle state. On the other hand, the energy gain due to the additional $\Lambda$ particle for the Hoyle state is only a half of that for the ground state, which derives from difference between the $\C(0_2^+)+\Lambda$ and $\C(0_1^+)+\Lambda$ folding potentials, reflecting the difference of density distributions of both states~\cite{Ya85}.

We should also mention that contrary to the situation in $\LC$, the excited states of $\C$ have been extensively studied by using the cluster models such as the RGM~\cite{Fu78}, Generator Coordinate Method (GCM)~\cite{Br66,Hi53,Ue77,De87}, and OCM~\cite{Ho74}. All these models could succeed in reproducing the experimental data for the low-lying states including the Hoyle state~\cite{Fu80}.
In particular, in the last 15 years, it was established that the Hoyle state has a ``Bose-condensate'' like feature, where the $3\alpha$ clusters weakly interact with each other like a gas and occupy an identical orbit~\cite{To01,Fu03,Ya04,Ma04,Ya05,Fu05,Fu06,Fu09,Fu15a,Sc16,To17,Fr17}. The important contribution to the establishment of this concept was given by the so-called Tohsaki-Horiuchi-Schuck-R\"opke (THSR) wave function of the $\alpha$ condensate type character~\cite{To01,Fu02}, in which the RGM and GCM wave functions for the Hoyle state were shown to be almost equivalent to a single configuration of the THSR wave function~\cite{Fu03}.

While it is now widely accepted that the Hoyle state has the new feature as $\alpha$ condensate, exploration of excited states above the Hoyle state is in a frontier of nuclear structure problems. For example, the second $J^\pi=2^+$ and $4^+$ states were very recently observed at around $10$ MeV~\cite{It04,Fr09,It11,Zi11,Zi13} and $13.3$ MeV~\cite{Fr11}, respectively. Their relation with the Hoyle state, whether or not they form a rotational band with the Hoyle state, has been strongly argued~\cite{Fu80,Bi00,La14,Fu15}. 

Furthermore, there has been known a $J^\pi=0^+$ state at $10.3$ MeV with about $3$ MeV width above the Hoyle state, but very recently it was reported in experiment that this broad $0^+$ state is decomposed into two peaks at $9.04$ MeV and $10.56$ MeV, with widths of $1.45$ MeV and $1.42$ MeV, corresponding to the $0_3^+$ and $0_4^+$ states, respectively~\cite{It11,It13}. Theoretically the $0_3^+$ and $0_4^+$ states were obtained around the $10$ MeV region by using the $3\alpha$ OCM, with a proper resonance boundary condition~\cite{Ku05,Oh13}.
In Ref.~\cite{Ku05}, it was mentioned that the $0_3^+$ has an $S$-wave dominant structure, according to an extrapolation method. On the other hand, the Antisymmetrized Molecular Dynamics (AMD) and Fermionic Molecular Dynamics (FMD) calculations predicted that the $0_4^+$  state has a bent-armed shape of the $3\alpha$ clusters, which resembles a linear-chain structure of $3\alpha$ clusters~\cite{Mo56,Su72,Su14}, though in these calculations the $0_3^+$ state is missing~\cite{En07b,Ch07,Su15}. 

Very recently these $0^+$ states above the Hoyle state are also investigated by using the THSR wave function with an extension to include $\Be+\alpha$ asymptotic configuration~\cite{Fu15,Fu16,Zh16} and a treatment of resonances~\cite{Fu05}, since this wave function is known to be highly reliable for the description of the Hoyle state~\cite{Fu03}. The both $0^+$ states are reproduced well, and it is concluded that the $0_4^+$ state predominantly has the linear-chain configuration composed of the $3\alpha$ clusters and the $0_3^+$ state is considered to be a family of the Hoyle state~\cite{Fu16}, in which the internal motions of the $3\alpha$ clusters experience a monopole excitation from the Hoyle state. 


The purpose of this study is to investigate $\LC$ to exhibit what happens when the $\Lambda$ particle is added to $\C$. Not only whether the Hoyle analog state appears or not, which has been explored in neighboring nuclei such as $\Ox$~\cite{Fu08,Fu10}, ${^{11}{\rm B}}$~\cite{Ka07,En07a,Ya10}, and ${^{13}{\rm C}}$~\cite{Ya08,Ya15}, but also what kind of structural change is expected, whether the linear-chain state in the $0_4^+$ state of $\C$ remains or not, and whether a new state appears or not, are of highly interest. By using the extended version of the THSR wave function, which can successfully describe the excited states of the core $\C$ nucleus, we discuss the similarities and differences between the structures of $\C$ and $\LC$. 

In the recent works on the $3\alpha$ cluster structures in $\C$, one of the authors (Y. F.) employed the following microscopic cluster model that is referred to as the THSR wave function~\cite{Zh13,Zh14,Fu15,Fu16}:
\begin{eqnarray}
&&\Phi^{\rm THSR}_{3\alpha}(\vc{\beta}_1,\vc{\beta}_2) \nonumber \\
&&\hspace{-0.25cm} = {\cal A}\Big[ \exp \Big\{\hspace{-0.1cm} - \hspace{-0.1cm} \sum_{i=1}^2 \mu_i \hspace{-0.2cm} \sum_{k=x,y,z}\hspace{-0.2cm} \frac{\xi_{ik}^2}{b^2+2\beta_{ik}^2}\Big\} \phi(\alpha_1)\phi(\alpha_2)\phi(\alpha_3) \Big], \label{eq:1}
\end{eqnarray}
with
\begin{equation}
\phi(\alpha_i) \propto \exp\Big[-\sum_{1\leq k<l \leq 4}({\vc r}_{i,k} - {\vc r}_{i,l})^2/(8b^2)\Big], \label{eq:int_alpha}
\end{equation}
where $\vc{\beta}_i$ $(i=1,2)$ are variational parameters to characterize a size and shape of the nucleus, while the other parameter $b$ is fixed at $1.348$ fm, which is close to the size of the $\alpha$ particle in free space, throughout this paper. $\vc{\xi}_i$ $(i=1,2)$ are the Jacobi coordinates between the $3\alpha$ clusters, $\vc{\xi}_1=\vc{R}_1-\vc{R}_2$ and $\vc{\xi}_2=\vc{R}_3-(\vc{R}_1+\vc{R}_2)/2$, where $\vc{R}_i=(\vc{r}_{4(i-1)+1}+\cdots+\vc{r}_{4(i-1)+4})/4$ $(i=1,2,3)$, and $\mu_i=2i/(i+1)$. ${\cal A}$ is the antisymmetrizer acting on 12 nucleons. We should mention that this model was employed in the previous study of $\C$ and is shown to reproduce very well the excited states above the Hoyle state, the Hoyle band $(0_2^+, 2_2^+,4_2^+)$, and $0_3^+$ and $0_4^+$ states~\cite{Fu15,Fu16}.

In the present work, we add one $\Lambda$ particle to this THSR wave function in the following way,
\begin{equation}
\Phi^{\rm H-THSR}_{3\alpha}(\vc{\beta}_1,\vc{\beta}_2,\vc{\kappa})
 ={\cal A}\Big[ \exp \Big\{\hspace{-0.1cm} - \hspace{-0.1cm} \sum_{i=1}^2 \mu_i \hspace{-0.2cm} \sum_{k=x,y,z}\hspace{-0.2cm} \frac{\xi_{ik}^2}{b^2+2\beta_{ik}^2}\Big\} \phi^3(\alpha) \Big]\varphi_{\Lambda}(\vc{\kappa}), \label{eq:2} 
\end{equation}
with 
\begin{equation}
\varphi_{\Lambda}(\vc{\kappa})=\exp\Big(-\mu_{\Lambda} \sum_{k=x,y,z}\frac{\xi_{3k}^2}{2b^2+\kappa_k^2}\Big),
\end{equation}
where the $\varphi_\Lambda(\vc{\kappa})$ is the wave function of the $\Lambda$ particle with its width parameter, $\vc{\kappa}$, which is out of the antisymmetrization, $\mu_\Lambda=12m_\Lambda/(12m_N+m_\Lambda)$ with $m_N$ and $m_\Lambda$ being nucleon and $\Lambda$-particle masses, respectively, and $\vc{\xi}_3=\vc{R}_4-(\vc{R}_1+\vc{R}_2+\vc{R}_3)/3$ with $\vc{R}_4$ a position vector of the $\Lambda$ particle. We refer to this wave function as the Hyper-THSR (H-THSR) wave function~\cite{Fu14} in this work.

We should note that the present model wave function breaks rotational symmetry for the center-of-mass motions of the $\alpha$ particles and $\Lambda$ particle, though in our previous work the $\Lambda$ particle motion is kept spherical~\cite{Fu14}. Throughout this work, axially symmetric deformation is taken into account, i.e. $\beta_x=\beta_y\neq \beta_z$ and $\kappa_x=\kappa_y\neq \kappa_z$. This allows for coupling of the $\Lambda$ particle in non-zero spin partial waves with the deformable $\C$ core. The rotational symmetry is restored by introducing the angular-momentum projection operator, like $\Phi_{JM}^{\rm H-THSR} = {\cal N} {\widehat P}^J_M \Phi^{\rm H-THSR}_{3\alpha}$ with ${\cal N}$ a normalization constant.
 Then we suppose the following linear combination in terms of the parameters, $\vc{\beta}_1$, $\vc{\beta}_2$, and $\vc{\kappa}$, to obtain the energy eigenstates:
\begin{equation}
\Psi^{(\lambda)}_{JM} = \sum_{\vc{\beta}_1,\vc{\beta}_2,\vc{\kappa}}f_\lambda(\vc{\beta}_1,\vc{\beta}_2,\vc{\kappa}) \Phi^{\rm H-THSR}_{JM}(\vc{\beta}_1,\vc{\beta}_2,\vc{\kappa}). \label{eq:cutoff2}
\end{equation}
The coefficients of the linear combination, $f_\lambda(\vc{\beta}_1,\vc{\beta}_2,\vc{\kappa})$ can be determined by solving the Hill-Wheeler equation as follows:
\begin{eqnarray}
&&\sum_{\vc{\beta}_1^\prime,\vc{\beta}_2^\prime,\vc{\kappa}^\prime} \big\langle \Phi_{JM}^{\rm H-THSR} (\vc{\beta}_1,\vc{\beta}_2,\vc{\kappa}) \big|{\hat H}-E_\lambda \nonumber \\
&&\hspace{1cm} \big|\Phi_{JM}^{\rm H-THSR} (\vc{\beta}_1^\prime,\vc{\beta}_2^\prime,\vc{\kappa}^\prime) \big\rangle   f_\lambda(\vc{\beta}_1^\prime,\vc{\beta}_2^\prime,\vc{\kappa}^\prime)=0. \label{eq:ghweq}
\end{eqnarray}
In this calculation, we adopt the following values of the parameter set: $(\beta_{1x}=\beta_{1y},\ \beta_{1z},\ \beta_{2x}=\beta_{2y},\ \beta_{2z})=$ $(0.75\times 1.5^{i-1},\ 0.75\times 1.5^{j-1},\ 0.75\times 1.5^{i-1},\ 0.75\times 1.5^{j-1}\ {\rm fm})$, $(1.5,\ 3.0,\ 0.75\times 1.5^{i-1},\ 0.75\times 1.5^{j-1}\ {\rm fm})$, with $i,j=1,\cdots,7$, and $(\kappa_x=\kappa_y,\ \kappa_z)=(0.75\times 1.5^{k-1},\ 0.75\times 1.5^{l-1}\ {\rm fm})$, with $k,l=1,\cdots,6$, with restrictions, $\beta_{1z}\geq \beta_{1x}=\beta_{1y}$, $\beta_{2z}\geq \beta_{2x}=\beta_{2y}$, $\kappa_{z}\geq \kappa_{x}=\kappa_{y}$, $\beta_{1x}\beta_{1y}\beta_{1z}\leq 450$, $\beta_{2x}\beta_{2y}\beta_{2z}\leq 450$, and $\kappa_{x}\kappa_{y}\kappa_{z}\leq 450$.

For Hamiltonian, we use the following microscopic one composed of the kinetic energy $T_i$, the effective nucleon-nucleon force, $V^{(NN)}_{ij}$, the Coulomb force $V_{ij}^{(C)}$, and $\Lambda N$ force $V_i^{\Lambda N}$:
\begin{equation}
H=\sum_{i=1}^{13}T_i - T_G  +\sum_{i<j}^{12}V_{ij}^{(C)} + \sum_{i<j}^{12} V^{(NN)}_{ij} + \sum_{i=1}^{12}V_{i}^{(\Lambda N)}, \label{eq:hml} 
\end{equation}
where $T_G$ is the spurious center-of-mass kinetic energy and the negligibly small $N\Lambda$ spin-orbit force is not taken into account in this work. For $V_{ij}^{(NN)}$ we adopt the same $NN$ force as used in Refs.~\cite{Fu15,Fu16}, Volkov No.~2 force~\cite{Vo65}, where the strength parameters are slightly modified~\cite{Fu80}. For the $N\Lambda$ interaction, we adopt the spin-independent parts of the YNG interaction, ESC04a~\cite{Ri06,Yam10}, where the fermi-momentum parameter is taken to be $k_F=1.0734\ {\rm fm}^{-1}$, for which the empirical values of $\Lambda$ binding energy of $\LH$ and $\LB$, i.e. $B_\Lambda(\LH)=3.12$ MeV and $B_\Lambda(\LB)=6.71$ MeV, respectively, are well reproduced by the present H-THSR ansatz.

In Fig.~\ref{fig:level}, the calculated $J^\pi=0^+$ spectra of $\C$ and $\LC$ and the observed $J^\pi=0^+$ spectrum of $\C$ are shown. 
The four $0^+$ states in $\C$ shown in this figure are also obtained by the THSR ansatz~\cite{Fu15,Fu16}: 
The $0_1^+$ state of $\C$ has a shell-model-like structure and the observed energy and r.m.s. radius are well reproduced~\cite{Zh14}. The $0_2^+$ state, i.e. the Hoyle state, is located very closely to the $3\alpha$ breakup threshold, and accordingly has the loosely coupled $3\alpha$ cluster structure like a gas. The $0_3^+$ state has a higher nodal structure in the internal motions of the $\alpha$ clusters than in the Hoyle state~\cite{Fu16}. In Ref.~\cite{Zh16}, this state is also discussed and is concluded to be the vibrational excitation from the Hoyle state. On the other hand, the $0_4^+$ state is shown to have a $3\alpha$ linear-chain component as a dominant configuration. Thus, while the $0_3^+$ state is considered to be a family of the Hoyle state, the $0_4^+$ state has a quite different structure from the $0_2^+$ and $0_3^+$ states.

In $\LC$, first we should mention that we simply denote the spectrum of $\LC$ as $J^\pi=0^+$ states, neglecting the intrinsic spin $1/2$ of the $\Lambda$ particle, for simplicity. From this figure, one can see that $\LC$ gives more decay channels than $\C$, i.e. $\Lambda+\C(0_1^+)$, $\alpha+\LB(0_1^+)$, $\Lambda+\C(2_1^+)$, $\alpha+\LB(2_1^+)$, $\Be(0^+)+\LH$, $\alpha+\LB(0_2^+)$, and $\Be(2^+)+\LH$. Up to the $3\alpha+\Lambda$ threshold, we have the four $J^\pi=0^+$ states, which appear to correspond to the four $J^\pi=0^+$ states in $\C$. We can see that the $0_2^+$ in $\LC$ is obtained as a bound state which is located below the lowest threshold, $\Lambda+\C(0_1^+)$, by $0.8$ MeV.
Although the $0_3^+$ and $0_4^+$ states are obtained below the $3\alpha+\Lambda$ threshold, the new decay thresholds open, as mentioned above, and hence they inherently exist as resonances. Then one might suspect that these resonance states are difficult to handle in the present bound state approximation, in which nevertheless we obtain them as quasi stable states. However, according to the energy gain for all these states, we can see the contraction of the size of them, as we will discuss in the next in Table~\ref{tab:rms}. This indicates that the amplitudes of the wave functions of those states are pushed toward the inside, which can play a role in preventing the states from decaying in a short life time, i.e. from having a broad decay width. 
Although we have checked the stability of the solutions in Eq.~(\ref{eq:ghweq}) by varying the adopted values of the parameter set, we further calculated the widths decaying into the possible channels, by using the separation energy method based on the $R$-matrix theory~\cite{La58}. The $0_3^+$ state is found to mainly decay into the $\Lambda+\C(0_1^+)$ and $\alpha+\LB(0_1^+)$ channels, whose partial decay widths are calculated to be at most $0.6$ MeV and $0.4$ MeV, respectively. Decay width into the other channel, $\Lambda+\C(2^+)$, is negligibly small. These values are much smaller than the corresponding decay energies, $3.1$ MeV and $2.6$ MeV, respectively. The $0_4^+$ state mainly decays into the $\alpha+\LB(0_1^+)$, $\alpha+\LB(2_1^+)$, and $\Be(0^+)+\LH$ channels with their partial widths, at most, $1.1$ MeV, $1.0$ MeV, and $0.4$ MeV, respectively, while the widths decaying into the other channels are negligibly small. The decay energies, $5.4$ MeV, $2.5$ MeV, and $1.8$ MeV, respectively, are smaller enough than the corresponding widths. This is different from the situation in the $0_3^+$ state of $\C$, whose decay energy and width are comparable, $1.77$ MeV and $1.45$ MeV, respectively~\cite{It11}, in which therefore some techniques to impose resonance boundary condition are necessary in theoretical treatment~\cite{Fu15,Ku05,Oh13,Fu06a}. These results thus allow us to safely discuss these states in $\LC$ within the bound state approximation. Actually the energy deviation of the $0_3^+$ and $0_4^+$ states against the variation of adopted mesh points to solve Eq.~(\ref{eq:ghweq}) is still within only a few hundred keV, while the convergence of the bound $0_1^+$ and $0_2^+$ states is in an order of ten keV.

We note here that also within the bound state approximation, we predict the existence of $\LB(0_2^+)$ state at $1.4$ MeV above the $\alpha+\LH$ threshold ($1.7$ MeV blow the $2\alpha+\Lambda$ threshold, see Fig.~\ref{fig:level}). Although this state has not yet been discussed before, our calculation by using the $2\alpha+\Lambda$ H-THSR wave function for $\LB$ gives the eigenstate at this energy position, as a very much stable solution against the variation of basis functions. Since this state is still located below the $2\alpha+\Lambda$ threshold and only the $\alpha+\LH$ decay channel is open, it must have an $\alpha+\LH$ resonant nature. In fact, our calculation of reduced width amplitude (RWA) of $\alpha+\LH$ channel shows a well developed $\alpha+\LH$ cluster structure with $3$ nodes (the ground state $\LB(0_1^+)$ has 2 nodes for the RWA).
We will discuss in more detail the resonance nature of this state in a forthcoming paper.


\begin{table}
\begin{center}
\caption{Root mean square radii of $\LC$ $(R_{\rm rms})$ and of the $\C$ core $(R_{\rm rms}^{({\rm c})})$, r.m.s. distance between the $\Lambda$ particle and $\C$ core, for the $0_1^+$ - $0_4^+$ states of $\LC$, and for comparison, r.m.s. radius of the $0_1^+$ - $0_4^+$ states in $\C$, are shown in a unit of fm. ${\cal E}_\Lambda$ values for the $0_1^+$ - $0_4^+$ states in $\LC$ shown in Fig.~\ref{fig:level}, which are defined as differences of binding energies of the states of $\LC$ from those of the corresponding states in $\C$, and excitation energies $E_{\rm exc}$, are also shown, together with the corresponding experimental data.}\label{tab:rms}
\begin{tabular}{cccccccl}
\hline\hline
 & \multicolumn{5}{c}{$\LC$} & \multicolumn{2}{c}{$\C$} \\
 & $R^{({\rm c})}_{\rm rms}$ & $R^{({\rm c}-\Lambda)}_{\rm rms}$ & $R_{\rm rms}$ & $E_{\rm exc}$ & ${\cal E}_\Lambda$ & \multicolumn{2}{c}{$R_{\rm rms}$} \\
\hline
\raisebox{-1.8ex}[0pt][0pt]{$0_1^+$} & \raisebox{-1.8ex}[0pt][0pt]{$2.2$} & \raisebox{-1.8ex}[0pt][0pt]{$2.1$} & \raisebox{-1.8ex}[0pt][0pt]{$2.2$} &  & $15.0$ & \multicolumn{2}{c}{\raisebox{-1.8ex}[0pt][0pt]{$2.4$}} \\
 &  &  &  &  & $({\rm exp}: 11.69)$ &  &  \\
$0_2^+$ & $2.8$ & $3.4$ & $2.9$ & $14.4$ & $8.5$ & \multicolumn{2}{c}{$3.7$} \\
$0_3^+$ & $3.1$ & $4.8$ & $3.2$ & $18.1$ & $8.3$ & \multicolumn{2}{c}{$4.2$ $(0_4^+)$} \\
$0_4^+$ & $4.3$ & $4.8$ & $4.3$ & $20.9$ & $4.1$ & \multicolumn{2}{c}{$4.7$ $(0_3^+)$} \\
\hline\hline
\end{tabular}
\end{center}
\end{table}

In Table~\ref{tab:rms}, we show the root mean square (r.m.s.) radius of the core $R_{\rm rms}^{({\rm c})}$, the r.m.s. distance between the core and $\Lambda$ particle $R_{\rm rms}^{({\rm c}-\Lambda)}$, and the r.m.s. radius of $\LC$, $R_{\rm rms}$, together with the r.m.s. radius of isolated $\C$ calculated in Ref.~\cite{Fu15}.
We can see that the higher excited state has a larger r.m.s. radius.
 The most important result that can be deduced from the calculated r.m.s. radii for these states is that only the $0_4^+$ state is qualified to be a gaslike cluster state, since only this state has a larger r.m.s. radius $4.3$ fm than the one of the Hoyle state $3.7$ fm, which is the typical gaslike cluster state of the $3\alpha$ clusters. The r.m.s. radii for the other states are small enough compared with the Hoyle state radius. We will later discuss the mechanism in detail.

\begin{figure}[htbp]
\begin{center}
\includegraphics[scale=0.80]{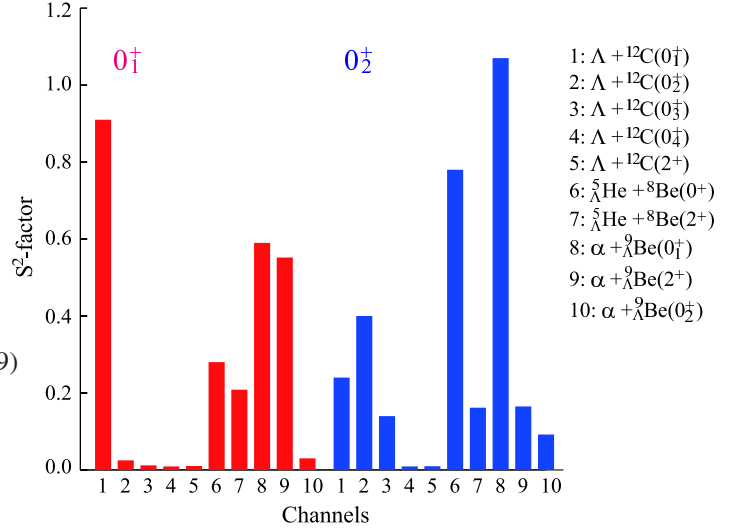}
\caption{(color online) $S^2$-factor of the $0_1^+$ and $0_2^+$ states for various channels, $\Lambda+\C(0_1^+,\ 0_2^+,\ 0_3^+,\ 0_4^+,\ {\rm and}\ 2^+)$, which are denoted as $1$, $2$, $3$, $4$, and $5$, respectively, and $\LH+\Be(0^+)$, $\LH+\Be(2^+)$, $\alpha+\LB(0_1^+)$, $\alpha+\LB(2_1^+)$, and $\alpha+\LB(0_2^+)$, which are denoted as $6$, $7$, $8$, $9$, and $10$, respectively.}
\label{fig:sfact1}
\end{center}
\end{figure}

\begin{figure}[htbp]
\begin{center}
\includegraphics[scale=0.80]{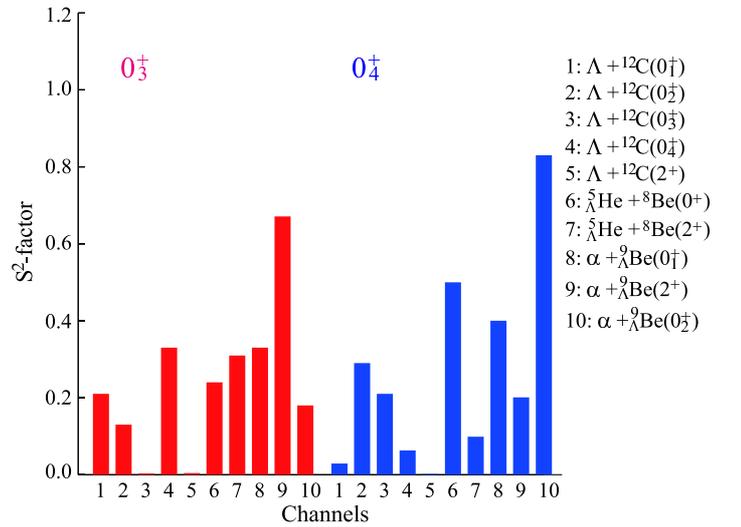}
\caption{(color online) $S^2$-factor of the $0_3^+$ and $0_4^+$ states for various channels, $\Lambda+\C(0_1^+,\ 0_2^+,\ 0_3^+,\ 0_4^+,\ {\rm and}\ 2^+)$, which are denoted as $1$, $2$, $3$, $4$, and $5$, respectively, and $\LH+\Be(0^+)$, $\LH+\Be(2^+)$, $\alpha+\LB(0_1^+)$, $\alpha+\LB(2_1^+)$, and $\alpha+\LB(0_2^+)$, which are denoted as $6$, $7$, $8$, $9$, and $10$, respectively.}
\label{fig:sfact2}
\end{center}
\end{figure}

We also show in Table~\ref{tab:rms} the $\Lambda$ binding energies, ${\cal E}_\Lambda$, which are defined as the binding energies for the $0_\lambda^+$ $(\lambda=1,\cdots,4)$ states relative to the corresponding $\C(0_\lambda^+)+\Lambda$ $(\lambda=1,\cdots,4)$ thresholds. Here we should note that concerning the correspondence between both the spectra in $\LC$ and $\C$, the counterparts of the $0_3^+$ and $0_4^+$ states in $\LC$ are most likely the $0_4^+$ and $0_3^+$ states in $\C$, respectively, as we indicate in Fig.~\ref{fig:level}. We will later explain why this assignment should be better and this inversion of level ordering happens, with analysis of $S^2$-factor in Figs.~\ref{fig:sfact1} and \ref{fig:sfact2}. Then the ${\cal E}_\Lambda$ values of the $0_1^+$, $0_2^+$, $0_3^+$, and $0_4^+$ states in $\LC$ are accordingly defined as the binding energies measured from the $\C(0_1^+)+\Lambda$, $\C(0_2^+)+\Lambda$, $\C(0_4^+)+\Lambda$, and $\C(0_3^+)+\Lambda$ states, respectively. We also show in Table~\ref{tab:rms} the experimental data of ${\cal E}_\Lambda$ for the ground state, and this value, $11.69$ MeV, is sizably smaller than the calculated one $15.0$ MeV. 
This is, however, well known to happen when using the $N\Lambda$ YNG potential, in which its density dependence is controlled by the adjustable fermi-momentum parameter $k_F$, since the present $k_F$ value adopted to reproduce the binding energy of $\LB$ with clustering nature, as well as that of $\LH$, is not appropriate for the ground state of $\LC$ with more compact shell-model-like structure~\cite{Hi97}.

As mentioned in Introduction, the ground state that has saturation density experiences almost no shrinkage by the additional $\Lambda$ particle but gains much larger binding energy than cluster states like the Hoyle state whose density is far below the saturation. That difference can be seen for the $0_1^+$ and $0_2^+$ states. With the $\Lambda$ particle injected, the density of the Hoyle state becomes $(3.7/2.8)^3=2.4$ times larger, though ${\cal E}_\Lambda=8.5$ MeV is only about a half of that of the ground state, ${\cal E}_\Lambda=15.0$ MeV. The $0_3^+$ state also behaves like the Hoyle state, i.e. $(4.2/3.1)^3=2.5$ times larger density and ${\cal E}_\Lambda=8.3$ MeV. However, further mechanism appears to work for the cluster states, considering the behavior of the $0_4^+$ state. For this state, the change of density and the energy gain, $(4.7/4.3)^3=1.3$ and ${\cal E}_\Lambda=4.1$ MeV, respectively, are only halves of those of the $0_2^+$ and $0_3^+$ states. The injected $\Lambda$ particle does not appear to play an efficient role as a glue and does not gain so much the binding energy for this state.  We then have to answer the question of what this quite a large difference comes from, as well as to explain the reason why the reversal assignment of both $0_3^+$ and $0_4^+$ states in $\C$ and $\LC$ is justified.

These questions can be answered with the analysis of $S^2$-factor for the four $0^+$ states decaying into various channels, defined as,
\begin{equation}
S^2_i = \int (r{\cal Y}_i(r) )^2 dr, \label{eq:s2}
\end{equation}
where ${\cal Y}_i(r)$, the reduced width amplitudes (RWAs), are defined below,
\begin{eqnarray}
\hspace{-0.75cm}
{\cal Y}_i(r)= \left\{
\begin{array}{l}
\displaystyle \sqrt{\frac{12!}{12!1!}} \Big\langle [\Psi_i(\C),Y_{00}({\widehat \xi}_i)]_{00}\frac{\delta(\xi_i-r)}{\xi_i^2}  \Big|\Psi^{(\lambda)}_{J=0}\Big\rangle  \\
  \hspace{5cm} (i=1,\cdots,5) \\
 \\
\displaystyle \sqrt{\frac{12!}{8!4!}}  \Big\langle [\Psi_i(\Be),Y_{00}({\widehat \xi}_i)]_{00}\frac{\delta(\xi_i-r)}{\xi_i^2}\Psi(\LH)  \Big|\Psi^{(\lambda)}_{J=0}\Big\rangle \\
 \hspace{5cm} (i=6, 7)   \\
 \\
\displaystyle \sqrt{\frac{12!}{8!4!}} \Big\langle [\Psi_i(\LB),Y_{00}({\widehat \xi}_i)]_{00}\frac{\delta(\xi_i-r)}{\xi_i^2} \Psi(\alpha) \Big|\Psi^{(\lambda)}_{J=0}\Big\rangle \\ 
 \hspace{5cm} (i=8,\cdots, 10). \label{eq:ov}
\end{array}
\right.
\end{eqnarray}
In the above equations, $\vc{\xi}_i$ are relative distances between $\C$ and $\Lambda$ particle for $i=1,\cdots,5$, $\Be$ and $\LH$ for $i=6,7$, and $\LB$ and $\alpha$ cluster for $i=8,\cdots,10$. $\Psi_i(\C)$ with $i=1,2,3,4$, and $5$ denote the wave functions of the $0_1^+$, $0_2^+$, $0_3^+$, $0_4^+$, and $2^+$ states of $\C$, respectively, calculated by using the THSR ansatz. $\Psi_{i}(\Be)$ with $i=6,7$ denote the wave functions of the $0^+$ and $2^+$ states of $\Be$, respectively, also calculated by the THSR ansatz. $\Psi_i(\LB)$ with $i=8$, $9$, and $10$ denote the wave functions of the $0_1^+$, $2^+$ and $0_2^+$ states of $\LB$, respectively, calculated by the present H-THSR ansatz.

In Figs.~\ref{fig:sfact1} and \ref{fig:sfact2}, the $S^2$-factors for the four $0^+$ states decaying into various channels (Fig.~\ref{fig:sfact1} for the $0_1^+$ and $0_2^+$ states and Fig.~\ref{fig:sfact2} for the $0_3^+$ and $0_4^+$ states) are shown. 
First we explain which $0^+$ states in $\C$ are the counterparts of the $0_1^+$ - $0_4^+$ states in $\LC$, by focusing on the $\Lambda+\C(0^+_i)$ channels with $i=1,\cdots,4$ only. The $0^+$ states in $\LC$ must have large contributions from the corresponding $\Lambda+ \C(0^+)$ channels. For example, in Fig.~\ref{fig:sfact1}, the $0_1^+$ state and $0_2^+$ state have the largest components in the $\Lambda+\C(0_1^+)$ and $\Lambda+\C(0_2^+)$ channels, respectively, within $i=1,\cdots,4$. This means that the $0_1^+$ and $0_2^+$ states in $\LC$ correspond to the $0_1^+$ and $0_2^+$ states in $\C$, respectively. However, for the $0_3^+$ state the largest contribution, within $i=1,\cdots,4$, is from the $\Lambda+\C(0_4^+)$ channel and almost nothing from the $\Lambda+\C(0_3^+)$ channel. On the contrary, the $0_4^+$ state has very little from $\Lambda+\C(0_4^+)$ channel and large $\Lambda+\C(0_3^+)$ component is seen, though $\Lambda+\C(0_2^+)$ component is slightly larger. We can therefore conclude that the $0_3^+$ and $0_4^+$ states in $\LC$ come from the $\Lambda$-particle coupling with the $0_4^+$ and $0_3^+$ states in $\C$, respectively. The fact that the $0_4^+$ state includes rather large $\Lambda+\C(0_2^+)$ component allows us to consider that this state is a gaslike cluster state.


Next let us try to answer the question of why the shrinkage and the energy gain for the $0_4^+$ state are much different from those for the $0_2^+$ and $0_3^+$ states. We should first investigate the nature of the $0_2^+$ state. From Fig.~\ref{fig:sfact1}, we can understand that the dominant component included in this state is $\alpha+\LB(0_1^+)$ configuration, though $\LH+\Be(0^+)$ configuration is also largely included. This is because the binding energy of the $\LB$ is so large that it is energetically favored that the injected $\Lambda$ particle shrinks the core and forms $\alpha+\LB(0_1^+)$ cluster configuration. In fact, this state is located at around the $\alpha+\LB(0_1^+)$ threshold as well as around the $\Lambda+\C(0_1^+)$ threshold (see Fig.~\ref{fig:level}), which follows the so-called threshold rule, indicating that cluster states appear around the corresponding threshold energy. It is to be noted that this state is overbound below the thresholds and appears as the bound state, since the shrinkage effect of the $\Lambda$ particle is strong enough to make this state very compact object. One can again see in Table~\ref{tab:rms} that the r.m.s. radius $2.9$ fm of this state is much smaller than that of the Hoyle state $3.7$ fm.


On the other hand, the $0_4^+$ state has the largest component of $\alpha+\LB(0_2^+)$, and in the second, $\Be(0^+)+\LH$ component, though a large contribution from the $\alpha+\LB(0_1^+)$ channel can still also be seen. Let us be reminded that the $\LB(0_2^+)$ state has a loosely coupled $\alpha+\LH$ structure.
This energy of the $\alpha+\LB(0_2^+)$ state is quite close to the position of the $0_4^+$ state of $\LC$, $1.6$ MeV higher than the $2\alpha+\LH$ threshold, so that this state must dominantly include this component, i.e. loosely coupled $2\alpha+\LH$ component. This situation gives us the answer of the question of why the $\Lambda$ particle cannot increase the density of this state so much and accordingly cannot gain enough the binding energy. For the $0_4^+$ state the $\Lambda$ particle couples with only one $\alpha$ cluster, to form $\LH$. The smaller overlap with nucleons reduces the shrinkage as well as the gain of the $\Lambda$ binding energy. In fact, this state appears around the $2\alpha+\LH$ threshold energy, for the threshold rule. The slightly higher energy position, $1.6$ MeV above the $2\alpha+\LH$ threshold, allows for the diluteness of this state.

The reason why this structure state emerges can be attributed to an orthogonality condition. Since the $0_2^+$ state largely has a configuration of $\alpha+\LB(0_1^+)$, where the $\Lambda$ particle moves inside the $\Be$ nucleus, the orthogonality to the $0_2^+$ state, which is satisfied by the $0_4^+$ state, prevents the $\Lambda$ particle from overlapping with the $\Be$ core ($2\alpha$ clusters). 
This orthogonality to the $0_2^+$ state with rather compact structure, as well as to the ground state, makes the $0_4^+$ state quite a large object, where the r.m.s. radius of the core amounts to $R_{\rm rms}^{({\rm c})}=4.3$ fm. As a result, the loosely-coupled $2\alpha+\LH$ cluster state, considered as an analog of the Hoyle state in $\C$, is built. We should mention that a configuration space used in the $0_3^+$ state is completely different from the ones used in the $0_4^+$ and $0_2^+$ states, as we will discuss later, so that, to good approximation, the configuration space used in the $0_3^+$ state satisfies the orthogonality to the ones used in the $0_2^+$ and $0_4^+$ states.


\begin{figure}[htbp]
\begin{center}
\includegraphics[scale=0.8]{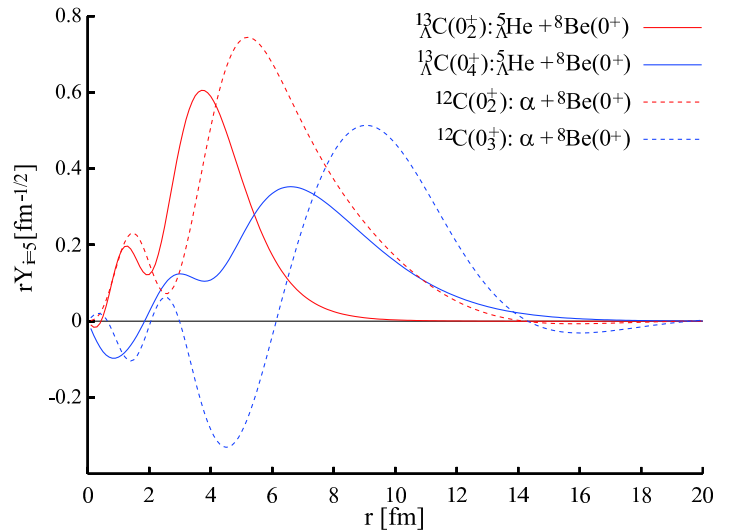}
\caption{(color online) RWAs of the $0_2^+$ (dotted curve in red) and $0_3^+$ (dotted curve in blue) state of $\C$ in the channel $\alpha+\Be(0^+)$ and of the $0_2^+$ (solid curve in red) and $0_4^+$ (solid curve in blue) states of $\LC$ in the channel $\LH+\Be(0^+)$ are shown.}
\label{fig:rwa3}
\end{center}
\end{figure}

We further investigate the structural change of the $0_4^+$ state in $\LC$ from the $0_3^+$ state in $\C$ by adding the $\Lambda$ particle. In Refs.~\cite{Fu15,Fu16} the author pointed out that the Hoyle state is further excited by strong monopole transition, to give rise to the $0_3^+$ state in $\C$, which has a higher nodal structure in the internal motions of the $3\alpha$ clusters than what the Hoyle state has. 

We show in Fig.~\ref{fig:rwa3} the RWAs defined in Eq.~(\ref{eq:ov}), $r{\cal Y}_{i=6}^{(\lambda)}(r)$, of the $0_2^+$ state (solid curve in red), with $\lambda=2$, and the $0_4^+$ state (solid curve in blue), with $\lambda=4$, in $\LC$ for the $\LH+\Be(0_1^+)$ channel, i.e. $i=6$, as a function of relative distance between $\LH$ and $\Be$. In comparison, the corresponding RWAs of the Hoyle state (dotted curve in red) and the $0_3^+$ state (dotted curve in blue) in $\C$ for the $\alpha+\Be(0_1^+)$ channel as a function of relative distance between $\alpha$ and $\Be$, are also shown.

We can see that the $0_3^+$ state in $\C$ (dotted curve in blue) has four nodes while in the Hoyle state (dotted curve in red) three nodes disappear and the corresponding oscillation remains. It is argued that these nodes appear as a result of the Pauli principle acting on the nucleons between the two clusters (in this case $\alpha$ and $\Be$) and outermost nodal point corresponds to the core radius, or the touching radius between the two clusters~\cite{Hi72}. Then the disappearance of the nodes indicates dissolution of the core and is the evidence that dilute gaslike cluster state appears, as long as a long tail behavior also remains, as in the Hoyle state~\cite{Ya05,Fu09,Fu15,Fu10}.

Let us now consider the RWAs for the $0_2^+$ (solid curve in red) and $0_4^+$ (solid curve in blue) states in $\LC$. While the both curves are dragged in inner region compared with those in $\C$, the most prominent feature is the disappearance of nodes for the $0_4^+$ state, though the innermost node only remains. This means that the $\Be$ core is dissolved in the $0_4^+$ state, completely unlike the case of the $0_3^+$ state in $\C$. We should also be aware that for the $0_4^+$ state a long tail is still very much developed, which is similar to the Hoyle state. This is also completely different from the $0_2^+$ state, in which the whole amplitude is pushed inside with no long tail any more. All these results again give strong support of our idea that the $0_4^+$ state has a gaslike structure of $2\alpha+\LH$ clusters, as a Hoyle analog state. 

It is important to mention that this Hoyle analog state uniquely appears in $\LC$ and we cannot expect in ${^{13}{\rm C}}$ $2\alpha+{^{5}{\rm He}}$ gas, since ${^{5}{\rm He}}$ is not bound~\cite{Ya15}, whereas $\LH$ is bound by $3.1$ MeV. 
Furthermore, in this Hoyle analog state, it is not energetically favored that, for example, the $\Lambda$ particle covalently exchanges with the gaslike $3\alpha$ clusters, rather than forming the $2\alpha+\LH$ gas, in which the potential energy gain due to its formation of $\LH$ cluster prevails against the loss of kinetic energy by its localization around an $\alpha$ cluster. This again raises the importance of the threshold rule.

We can also say that the fact that the $0_4^+$ state, which can be connected to the $0_3^+$ in $\C$, plays a role as the Hoyle analog state implies that the Hoyle state and the $0_3^+$ in $\C$ are intimately related to each other, like a family, as proposed in Refs.~\cite{Fu15,Fu16,Zh16}. 


\begin{figure}[htbp]
\begin{center}
\includegraphics[scale=0.8]{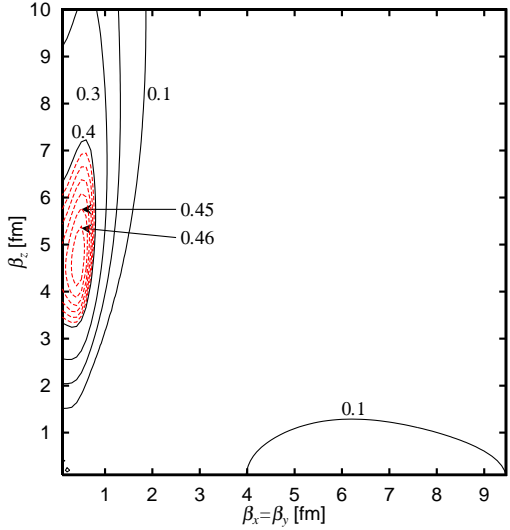}
\caption{(color online) Contour map of the squared overlap ${\cal O}_{\lambda=3}(\vc{\beta}_1=\vc{\beta}_2)$ in Eq.~(\ref{eq:ovlp}) for the $0_3^+$ state, in two parameter space, $\vc{\beta}_1=\vc{\beta}_2=(\beta_x=\beta_y,\ \beta_z)$. Black solid curves are drawn in a step of $0.1$ and red dotted curves, which cover the region of ${\cal O}_{\lambda=3}(\vc{\beta}_1=\vc{\beta}_2)\ge 0.41$, are in a step of $0.01$.}
\label{fig:lcs}
\end{center}
\end{figure}

Next we discuss the structure of the $0_3^+$ state. Since this state can be regarded as $\Lambda$ coupling to $\C(0_4^+)$ state that dominantly has a linear-chain configuration of the $3\alpha$ clusters, we can expect that such an exotic structure like the linear-chain is also realized in hypernuclei. 
According to Refs.~\cite{Fu80,Ku05,En07b,Fu15}, the $0_4^+$ state in $\C$ with the linear-chain-like configuration dominantly has $\alpha+\Be(2^+)$ component. From Fig.~\ref{fig:sfact2}, we can see that the $0_3^+$ state in $\LC$ also has the similar nature to the $0_4^+$ state in $\C$. The $0_3^+$ state in $\LC$  has the largest component of $\alpha+\LB(2^+)$ configuration, and sizable $\LH+\Be(2^+)$ component is also included, though the latter component appears to be suppressed since the energy of this channel is much higher than the position of the $0_3^+$ state (see Fig.~\ref{fig:level}). These results suggest that the $0_3^+$ state in $\LC$ considerably takes over the feature of the $0_4^+$ state in $\C$.

In order to further clarify this aspect for the $0_3^+$ state, we calculate the following squared overlap with the single THSR wave function, 
\begin{equation}
{\cal O}_{\lambda=3}(\vc{\beta}_1=\vc{\beta}_2) =\max_{\vc{\kappa}} |\langle \Phi_{J=0}^{\rm H-THSR}(\vc{\beta}_1=\vc{\beta}_2,\vc{\kappa}) | \Psi^{(\lambda=3)}_{J=0} \rangle |^2, \label{eq:ovlp}
\end{equation}
where the $\vc{\kappa}$ values are taken to give maximum values of the squared overlap for a given $\vc{\beta}_1=\vc{\beta}_2$ parameter value.

In Fig.~\ref{fig:lcs}, the contour map of the squared overlap Eq.~(\ref{eq:ovlp}) is shown. This figure corresponds to Fig.~6 in the previous publication~\cite{Fu16}, in which essentially the same quantity was calculated for the $0_4^+$ state in $\C$, i.e. for the linear-chain state in $\C$. We can see that, in principle, the behavior does not change. Only the parameter values $\beta_x=\beta_y,\ \beta_z$ giving the maximum shift from $(\beta_x=\beta_y,\ \beta_z)=(0.6,\ 6.7\ {\rm fm})$ to $(\beta_x=\beta_y,\ \beta_z)=(0.4,\ 4.6\ {\rm fm})$. This rather large shift along $z$-axix is due to the shrinkage effect by the $\Lambda$ particle. The optimal parameter value for the $\Lambda$ particle is then calculated $(\kappa_x=\kappa_y,\ \kappa_z)=(1.7,\ 4.8\ {\rm fm})$, indicating that the $\Lambda$ particle also keeps strongly prolately-deformed shape so as to cover the whole region occupied by the $3\alpha$ clusters. This helps to gain overlap between the $\Lambda$ particle and the $3\alpha$ clusters, and hence to gain the $\Lambda$ binding energy.

\begin{figure}[htbp]
\begin{center}
\includegraphics[scale=0.64, angle=270]{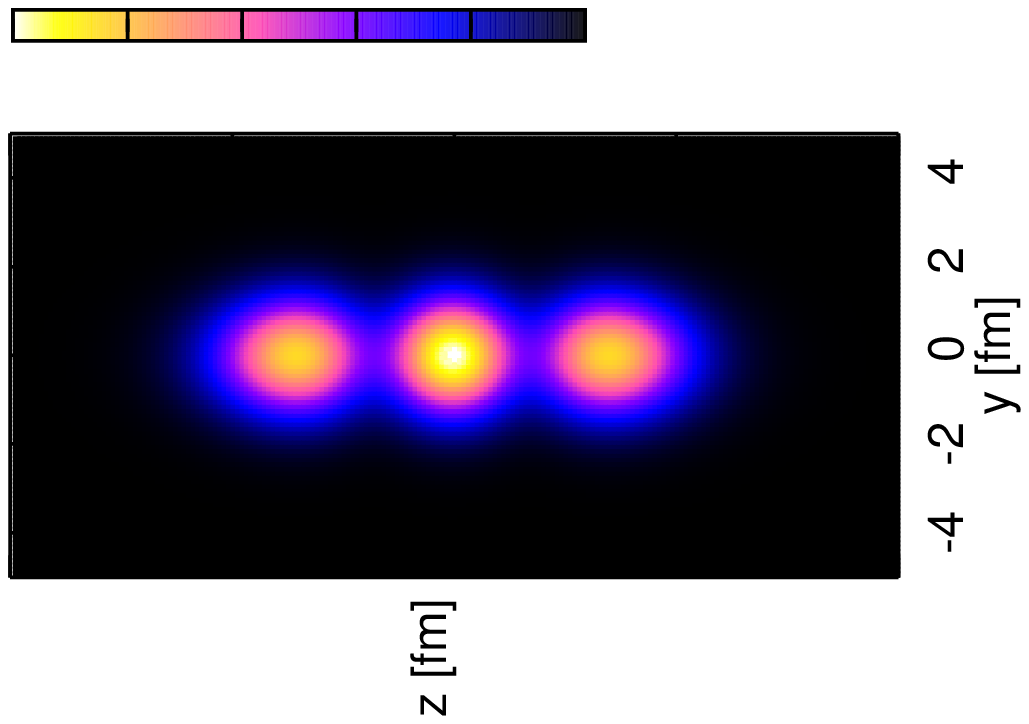}
\includegraphics[scale=0.64, angle=270]{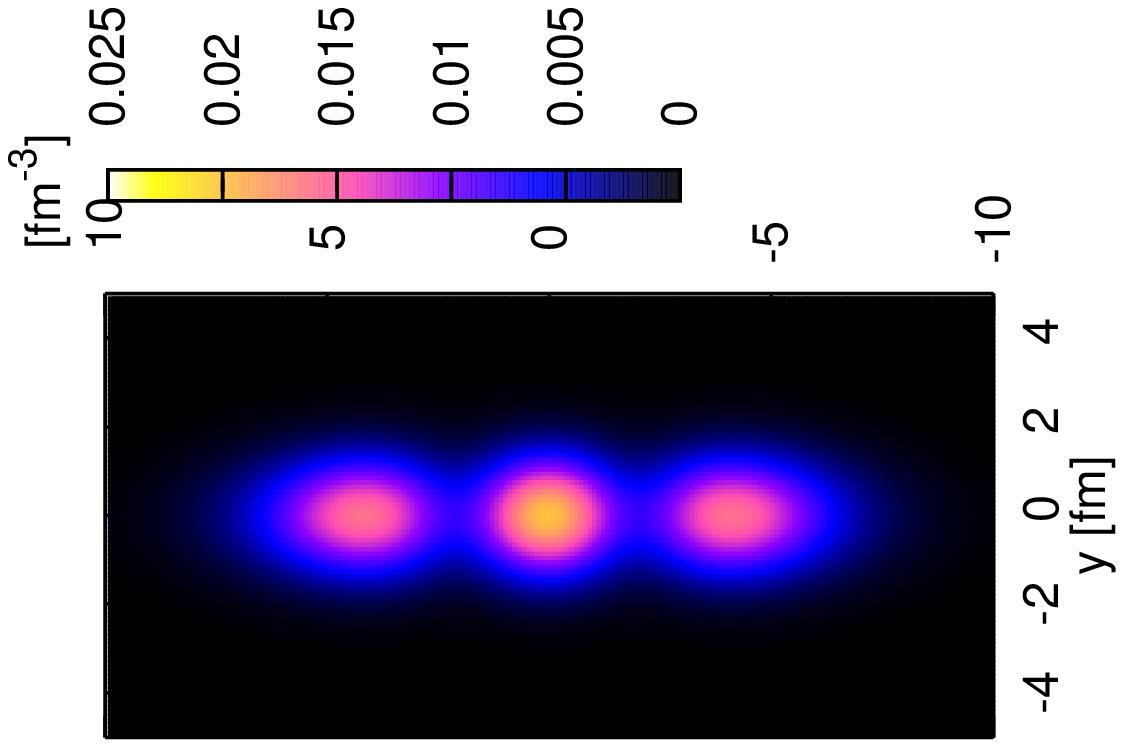}
\caption{(color online) (Left): Intrinsic density profile generated from the THSR wave function before angular-momentum projection, with $\vc{\beta}_1=\vc{\beta}_2=(\beta_x=\beta_y,\ \beta_z)=(0.4,\ 4.6\ {\rm fm})$ and $(\kappa_x=\kappa_y,\ \kappa_z)=(1.7,\ 4.8\ {\rm fm})$, which gives the maximal squared overlap, $0.47$, in Fig.~\ref{fig:lcs}. (Right): in comparison with the left, intrinsic density profile for the $0_4^+$ state in $\C$, which is taken from Fig.~7 in Ref.~\cite{Fu16}, is also shown (see Ref.~\cite{Fu16} for details).}
\label{fig:dsty}
\end{center}
\end{figure}

We also show in Fig.~\ref{fig:dsty} (Left) the intrinsic density profile of nucleons at the optimal parameter value $(\beta_x=\beta_y,\ \beta_z)=(0.4,\ 4.6\ {\rm fm})$, cut at $yz$-plane. In comparison, the one of the $0_4^+$ state in $\C$ is also shown at right, which is taken from Fig.~7 in Ref.~\cite{Fu16} and is calculated at the optimal parameter value $(\beta_x=\beta_y,\ \beta_z)=(0.6,\ 6,7\ {\rm fm})$ in the same THSR ansatz. The distinct linear-chain structure of the $3\alpha$ clusters can be seen. This component is included in the $0_3^+$ state by $47$ \% as a main configuration, which is almost the same as in the $0_4^+$ state of $\C$, $47$ \%. The nucleons are contracted along the $z$-direction from $\beta_z=6.7$ fm for $\C$ to $\beta_z=4.6$ fm for $\LC$, which we can also clearly see from both the figures. We can thus conclude that the $\Lambda$ particle makes the linear-chain structure in the $0_4^+$ state of $\C$ kept as it is and only shrinks the $3\alpha$ clusters along $z$-axis.

In conclusion, we investigated $J^\pi=0^+$ states in $\LC$ by using the THSR-type wave function, which very successfully reproduces the Hoyle state and the other two $J^\pi=0^+$ states in $\C$ recently observed above the Hoyle state. We obtained the four $0^+$ states in $\LC$, corresponding to the four $0^+$ states in $\C$. We showed that the coupling of the $\Lambda$ particle with the Hoyle state forces the $\LC(0_2^+)$ state to become a compact object with the $\alpha+\LB(0_1^+)$ structure and largely to lose its nature as a gas of the $\alpha$ clusters. Instead, as a result of the orthogonalization to the $\alpha+\LB(0_1^+)$ configuration, $\LC(0_4^+)$ state plays a role as the Hoyle analog state, in which gaslike $2\alpha+\LH$ cluster structure is formed. 
We also predict the existence of the linear-chain state in $\LC$, as the $0_3^+$ state, which is considered to be the $\Lambda$ particle coupling to the $0_4^+$ state in $\C$. 
It is remarkable that all these states appear around the corresponding cluster-breakup thresholds, according to the threshold rule, which is an important basis to describe cluster dynamics.
Although the $\LC(0_3^+)$ and $\LC(0_4^+)$ states are located above the lowest particle-decay threshold as resonances, we hope that this study will trigger interest of experimentalists and in near future all these excited states, including the $\LC(0_2^+)$ state, are observed in experiment.

\section*{Acknowledgements}
This work was partially performed with the financial support by HPCI Strategic Program of Japanese MEXT, JSPS KAKENHI Grant Number JP25400288, JP26400283, and RIKEN Incentive Research Projects.




\begin{thebibliography}{00}

\bibitem{Wi77}
K. Wildermuth and Y. C. Tang, {\it A Unified Theory of the Nucleus} (Vieweg, Braunschweig, 1977).
\bibitem{Ho12}
H. Horiuchi, K. Ikeda and K. Kat\={o}, Prog. Theor. Phys. Suppl. 192 (2012) 1.
\bibitem{Ik68}
K. Ikeda, N. Takigawa and H. Horiuchi, Prog. Theor. Phys. Suppl. Extra Num. 464 (1968) 1.
\bibitem{Zh17}
B. Zhou, Y. Funaki, H. Horiuchi and A. Tohsaki, to be submitted to Front. Phys. (2017).
\bibitem{Hi09}
E. Hiyama and T. Yamada, Prog. Part. Nucl. Phys. 63 (2009) 339; and references therein.
\bibitem{Ho54}
F. Hoyle, Astrophys. J. Suppl. Ser. 1 (1954) 121.
\bibitem{Co57}
C. W. Cook et al., Phys. Rev. 107 (1957) 508.
\bibitem{Fy11}
H.O.U. Fynbo and M. Freer, Physics 4 (2011) 94; and references therein.
\bibitem{Ya85}
T. Yamada, T. Motoba, K. Ikeda and H. Band\={o}, Prog. Theor. Phys. Suppl. 81 (1985) 104.
\bibitem{Hi97}
E. Hiyama, M. Kamimura, T. Motoba, T. Yamada, and Y. Yamamoto, Prog. Theor. Phys. 97 (1997) 881.
\bibitem{MBI83}
T. Motoba, H. Band\={o}, K. Ikeda, Prog. Theor. Phys. 70 (1983) 189; 71 (1984) 222.
\bibitem{Ya84}
T. Yamada, K. Ikeda, H. Band\={o}, and T. Motoba, Prog. Theor. Phys. 71 (1984) 985.
\bibitem{YMB86} 
Y. You-Wen, T. Motoba and H. Band\={o}, Prog. Theor. Phys. 76 (1986) 861.
\bibitem{Hi99}
E. Hiyama, M. Kamimura, K. Miyazaki, and T. Motoba, Phys. Rev. C. 59 (1999) 2351.
\bibitem{Is11}
M. Isaka, M. Kimura, A. Dot\'e, and A. Ohnishi, Phys. Rev. C 83 (2011) 044323; (2011) 054304.
\bibitem{Ba83} 
H. Band\={o}, K. Ikeda, and T. Motoba, Prog. Theor. Phys. 69 (1983) 918.
\bibitem{Ya85a}  
T. Yamada, K. Ikeda, H. Band\={o}, and T. Motoba, Prog. Theor. Phys. 73 (1985) 397; Phys. Rev. C 38 (1988) 854.
\bibitem{Wh37}
J. A. Wheeler, Phys. Rev. 52 (1937) 1083; 52 (1937) 1107.
\bibitem{Sa68}
S. Saito, Prog. Theor. Phys. 40 (1968) 893; 41 (1969) 705.
\bibitem{Hi00}
E. Hiyama, M. Kamimura, T. Motoba, T. Yamada, and Y. Yamamoto, Phys. Rev. Lett. 85 (2000) 270.
\bibitem{Fu08a}
Y. Funaki, H. Horiuchi, G. R\"opke, P. Schuck, A. Tohsaki, and T. Yamada, Phys. Rev. C 77 (2008) 064312.
\bibitem{Fu78}
Y. Fukushima and M. Kamimura, J. Phys. Soc. Jpn. Suppl. 44 (1978) 225; M. Kamimura, Nucl. Phys. A 351 (1981) r456.
\bibitem{Br66}
D. M. Brink, in Proceedings of the International School of Physics ``Enrico Fermi'', Course 36, edited by C. Bloch (Academic, New York/London, 1966) 247.
\bibitem{Hi53}
D. L. Hill, J. A. Wheeler, Phys. Rev. 89 (1953) 1120.
\bibitem{Ue77}
E. Uegaki, S. Okabe, Y. Abe, and H. Tanaka, Prog. Theor. Phys. 57 (1977) 1262; E. Uegaki, Y. Abe, S. Okabe, and H. Tanaka, {\it ibid.} 62 (1979) 1621.
\bibitem{De87}
P. Descouvemont and D. Baye, Phys. Rev. C 36 (1987) 54.
\bibitem{Ho74}
H. Horiuchi, Prog. Theor. Phys. 51 (1974) 1266; 53 (1975) 447.
\bibitem{Fu80}
Y. Fujiwara, H. Horiuchi, K. Ikeda, M. Kamimura, K. Kat\={o}, Y. Suzuki, and E. Uegaki, Prog. Theor. Phys. Suppl. 68 (1980) 29.
\bibitem{To01}
A. Tohsaki, H. Horiuchi, P. Schuck, and G. R\"opke, Phys. Rev. Lett. 87 (2001) 192501.
\bibitem{Fu03}
Y. Funaki, A. Tohsaki, H. Horiuchi, P. Schuck, and G. R\"opke, Phys. Rev. C 67 (2003) 051306(R).
\bibitem{Ya04}
T. Yamada, P. Schuck, Phys. Rev. C 69 (2004) 024309.
\bibitem{Ma04}
H. Matsumura, Y. Suzuki, Nucl. Phys. A 739 (2004) 238.
\bibitem{Ya05}
T. Yamada and P. Schuck, Eur. Phys. J A 26 (2005) 185.
\bibitem{Fu05}
Y. Funaki, A. Tohsaki, H. Horiuchi, P. Schuck, G. R\"opke, Eur. Phys. J. A 24 (2005) 321.
\bibitem{Fu06}
Y. Funaki, A. Tohsaki, H. Horiuchi, P. Schuck, G. R\"opke, Eur. Phys. J. A 28 (2006) 259.
\bibitem{Fu09}
Y. Funaki, H. Horiuchi, W. von Oertzen, G. R\"{o}pke, P. Schuck, A. Tohsaki, and T. Yamada, 
Phys. Rev. C 80 (2009) 064326.
\bibitem{Fu15a}
Y. Funaki, H. Horiuchi, A. Tohsaki, Prog. Part. Nucl. Phys. 82 (2015) 78.
\bibitem{Sc16}
P. Schuck, Y. Funaki, H. Horiuchi, G. R\"opke, A. Tohsaki, and T. Yamada, Phys. Scr. 91 (2016) 123001.
\bibitem{To17}
A. Tohsaki, H. Horiuchi, P. Schuck, and G. R\"opke, Rev. Mod. Phys. 89 (2017) 011002.
\bibitem{Fr17}
M.~Freer, H.~Horiuchi, Y.~Kanada-En'yo, D.~Lee and U.-G.~Mei{\ss}ner,
arXiv:1705.06192 [nucl-th].
\bibitem{Fu02}
Y. Funaki, H. Horiuchi, A. Tohsaki, P. Schuck, and G. R\"opke, Prog. Theor. Phys. 108 (2002) 297.
\bibitem{It04}
M. Itoh et al., Nucl. Phys. A 738 (2004) 268.
\bibitem{Fr09}
M. Freer et al., Phys. Rev. C 80 (2009) 041303(R).
\bibitem{It11}
M. Itoh et al., Phys. Rev. C 84 (2011) 054308.
\bibitem{Zi11}
W. R. Zimmerman, N. E. Destefano, M. Freer, M. Gai, and F. D. Smit, Phys. Rev. C 84 (2011) 027304.
\bibitem{Zi13}
W. R. Zimmerman, et al., Phys. Rev. Lett 110 (2013) 152502.
\bibitem{Fr11}
M. Freer et al., Phys. Rev. C 83 (2011) 034314.
\bibitem{Bi00}
R. Bijker and F. Iachello, Phys. Rev. C 61 (2000) 067305; Ann. Phys. (Amsterdam) 298 (2002) 334.
\bibitem{La14}
D. J. Mar\'in-L\'ambarri, R. Bijker, Ml Freer, M. Gai, Tz. Kokalova, D. J. parker, and C. Wheldon, Phys. Rev. Lett. 113 (2014) 012502.
\bibitem{Fu15}
Y. Funaki, Phys. Rev. C 92 (2015) 021302(R).
\bibitem{It13}
M. Itoh et al., J. Phys.: Conf. Ser. 436 (2013) 012006.
\bibitem{Ku05}
C. Kurokawa and K. Kat\={o}, Phys. Rev. C 71 (2005) 021301; Nucl. Phys. A 792 (2007) 87.
\bibitem{Oh13}
S. Ohtsubo, Y. Fukushima, M. Kamimura, and E. Hiyama, Prog. Theor. Exp. Phys. (2013) 073D02.
\bibitem{Mo56}
H. Morinaga, Phys. Rev. 101 (1956) 254; Phys. Lett. 21 (1966) 78.
\bibitem{Su72}
Y. Suzuki, H. Horiuchi and K. Ikeda, Prog. Theor. Phys. 47 (1972) 1517.
\bibitem{Su14}
T. Suhara, Y. Funaki, B. Zhou, H. Horiuchi, and A. Tohsaki, Phys. Rev. Lett. 112 (2014) 062501.
\bibitem{En07b}
Y. Kanada-En'yo, Prog. Theor. Phys. 117 (2007) 655.
\bibitem{Ch07}
M. Chernykh, H. Feldmeier, T. Neff, P. von Neumann-Cosel, and A. Richter, Phys. Rev. Lett. 98 (2007) 032501.
\bibitem{Su15}
T. Suhara and Y. Kanada-En'yo, Phys. Rev. C 91 (2015) 024315.
\bibitem{Fu16}
Y. Funaki, Phys. Rev. C 94 (2016) 024344.
\bibitem{Zh16}
B. Zhou, A. Tohsaki, H. Horiuchi, Z. Z. Ren, Phys. Rev. C 94 (2016) 044319.
\bibitem{Fu08}
Y. Funaki, T. Yamada, H. Horiuchi, G. R\"opke, P. Schuck, A. Tohsaki, Phys. Rev. Lett 101 (2008) 082502.
\bibitem{Fu10}
Y. Funaki, T. Yamada, A. Tohsaki, H. Horiuchi, G. R\"opke, and P. Schuck, Phys. Rev. C 82 (2010) 024312.
\bibitem{Ka07}
T. Kawabata et al., Phys. Lett. B 646 (2007) 6.
\bibitem{En07a}
Y. Kanada-En'yo, Phys. Rev. C 75 (2007) 024302.
\bibitem{Ya10}
T. Yamada and Y. Funaki, Phys. Rev. C 82 (2010) 064315.
\bibitem{Ya08}
T. Yamada and Y. Funaki, Int. J. Mod. Phys. E 17 (2008) 2101.
\bibitem{Ya15}
T. Yamada and Y. Funaki, Phys. Rev. C 92 (2015) 034326.
\bibitem{Zh13}
B. Zhou, Y. Funaki, H. Horiuchi, Z. Z. Ren, G. R\"opke, P. Schuck, A. Tohsaki, C. Xu, and T. Yamada, Phys. Rev. Lett. 110 (2013) 262501. 
\bibitem{Zh14}
B. Zhou, Y. Funaki, A. Tohsaki, H. Horiuchi, and Z. Z. Ren, Prog. Theor. Exp. Phys. (2014) 101D01.
\bibitem{Fu14}
Y. Funaki, T. Yamada, E. Hiyama, B. Zhou, and K. Ikeda, Prog. Theor. Exp. Phys. (2014) 113D01.
\bibitem{Vo65}
A. B. Volkov, Nucl. Phys. A 74 (1965) 33.
\bibitem{Ri06}
Th. A. Rijken, Phys. Rev. C 73 (2006) 044007.
\bibitem{Yam10}
Y. Yamamoto, T. Motoba, and Th. A. Rijken, Prog. Theor. Phys. Suppl. 185 (2010) 72.
\bibitem{La58}
A. M. Lane and R. G. Thomas, Rev. Mod. Phys. 30 (1958) 257.
\bibitem{Fu06a}
Y. Funaki, H. Horiuchi, and A. Tohsaki, Prog. Theor. Phys. 115 (2006) 115.
\bibitem{Hi72}
J. Hiura and R. Tamagaki, Prog. Theor. Phys. Suppl. 52 (1972) 25.
\end{thebibliography}


\end{document}